\documentclass{article} %
\usepackage{iclr2027_conference,times}

\usepackage{amsmath,amsfonts,bm}

\def\eqref#1{equation~\ref{#1}}

\def\1{\bm{1}}

\DeclareMathAlphabet{\mathsfit}{\encodingdefault}{\sfdefault}{m}{sl}
\SetMathAlphabet{\mathsfit}{bold}{\encodingdefault}{\sfdefault}{bx}{n}

\usepackage[T1]{fontenc}

\usepackage{hyperref}
\usepackage{url}

\title{The Optimal Ranking Tradeoffs\\ Between Two Ranking Scores\\ in Two-Class Classification}

\author{Sébastien Pi\'erard,
    Adrien Deli\`ege
    \& Marc Van Droogenbroeck\\
Montefiore Institute, University of Li\`ege, Li\`ege, Belgium\\
\texttt{\{S.Pierard, Adrien.Deliege, M.VanDroogenbroeck\}@uliege.be}
}

\usepackage{xargs}
\usepackage{fancybox}

\newcommand{\emphScore}[1]{{\color{teal}#1}}
\newcommand{\emphTradeoff}[1]{{\underline{\color{violet}#1}}}

\newcommand{\ie}{\emph{i.e.}\xspace}
\newcommand{\eg}{\emph{e.g.}\xspace}
\newcommand{\aka}{\emph{a.k.a.}\xspace}
\newcommand{\wrt}{\emph{w.r.t.}\xspace}
\newcommand{\iif}{\emph{iif}\xspace}
\newcommand{\iid}{\emph{i.i.d.}\xspace}

\newcommand{\first}{1\textsuperscript{st}\xspace}
\newcommand{\second}{2\textsuperscript{nd}\xspace}
\newcommand{\third}{3\textsuperscript{rd}\xspace}

\usepackage{multicol}
\usepackage{nicefrac}
\usepackage{amsthm}
\usepackage{pifont}
\usepackage{MnSymbol}
\usepackage{tikz}
\usepackage{xspace}
\usepackage{cleveref}
\usepackage{subfig}

\usepackage{thmtools}

\newtheorem*{question*}{Question}

\usepackage[dvipsnames]{xcolor}

\usepackage{soul}

\ExplSyntaxOn
\clist_const:Nn \l_soul_protect_clist { cite , citep , ref , eqref }
\NewDocumentCommand{\robusthl}{m}{
  \tl_set:Nn \l_tmpa_tl { #1 }
  \clist_map_inline:Nn \l_soul_protect_clist {
    \regex_replace_all:nnN { (\c{##1}[^{}]*\{[^{}]*\}) } %
                           { \c{mbox}\{\1\} } \l_tmpa_tl
  }
  \hl\l_tmpa_tl
}
\ExplSyntaxOff

\usepackage{titletoc}

\iclrfinalcopy %
\begin{document}

\global\long\def\sampleSpace{\Omega}%
\global\long\def\aSample{\omega}%
\global\long\def\eventSpace{\Sigma}%
\global\long\def\anEvent{E}%
\global\long\def\measurableSpace{(\sampleSpace,\eventSpace)}%
\global\long\def\expectedValueSymbol{\mathbf{E}}%

\global\long\def\aPerformance{P}%
\global\long\def\allPerformances{\mathbb{\aPerformance}_{\measurableSpace}}%
\global\long\def\aSetOfPerformances{\Pi}%
\global\long\def\randVarSatisfaction{S}%
\global\long\def\aScore{X}%
\global\long\def\allScores{\mathbb{\aScore}_{\measurableSpace}}%
\newcommandx\domainOfScore[1][usedefault, addprefix=\global, 1=\aScore]{\mathrm{dom}(#1)}%
\global\long\def\evaluation{\mathrm{eval}}%
\global\long\def\opFilter{\mathrm{filter}_\randVarImportance}%
\global\long\def\opNoSkill{\mathrm{no\text{\textendash{}}skill}}%
\global\long\def\opPriorShift{\mathrm{shift}_{\pi\rightarrow\pi'}}%
\global\long\def\opChangePredictedClass{\mathrm{change}_{\randVarPredictedClass}}%
\global\long\def\opChangeGroundtruthClass{\mathrm{change}_{\randVarGroundtruthClass}}%
\global\long\def\opSwapGroundtruthAndPredictedClasses{\mathrm{swap}_{\randVarGroundtruthClass\leftrightarrow\randVarPredictedClass}}%
\global\long\def\opSwapClasses{\mathrm{swap}_{\classNeg\leftrightarrow\classPos}}%
\global\long\def\allWorstPerformances{\frownie}
\global\long\def\allBestPerformances{\smiley}

\global\long\def\randVarGroundtruthClass{Y}%
\global\long\def\randVarPredictedClass{\hat{Y}}%
\global\long\def\allClasses{\mathbb{C}}%
\global\long\def\aClass{c}%
\global\long\def\classNeg{c_-}%
\global\long\def\classPos{c_+}%
\global\long\def\sampleTN{tn}%
\global\long\def\sampleFP{fp}%
\global\long\def\sampleFN{fn}%
\global\long\def\sampleTP{tp}%
\global\long\def\eventTN{\{\sampleTN\}}%
\global\long\def\eventFP{\{\sampleFP\}}%
\global\long\def\eventFN{\{\sampleFN\}}%
\global\long\def\eventTP{\{\sampleTP\}}%
\global\long\def\scorePTN{PTN}%
\global\long\def\scorePFP{PFP}%
\global\long\def\scorePFN{PFN}%
\global\long\def\scorePTP{PTP}%
\global\long\def\scoreAccuracy{A}%
\global\long\def\scoreExpectedSatisfaction{\aScore_{\randVarSatisfaction}}%
\global\long\def\scoreTNR{TNR}%
\global\long\def\scoreFPR{FPR}%
\global\long\def\scoreTPR{TPR}%
\global\long\def\scoreFNR{FNR}%
\global\long\def\scoreNPV{NPV}%
\global\long\def\scoreFOR{FOR}%
\global\long\def\scorePPV{PPV}%
\global\long\def\scorePrecision{\scorePPV}%
\global\long\def\scoreFDR{FDR}%
\global\long\def\scoreJaccardNeg{J_-}%
\global\long\def\scoreJaccardPos{J_+}%
\global\long\def\scoreCohenKappa{\kappa}%
\global\long\def\scoreScottPi{\pi}%
\global\long\def\scoreFleissKappa{\kappa}%
\global\long\def\scoreBalancedAccuracy{BA}%
\global\long\def\scoreWeightedAccuracy{WA}%
\global\long\def\scoreYoudenJ{J_Y}
\global\long\def\scorePLR{PLR}%
\global\long\def\scoreNLR{NLR}%
\global\long\def\scoreOR{OR}%
\global\long\def\scoreSNPV{SNPV}%
\global\long\def\scoreSPPV{SPPV}%
\global\long\def\scoreACP{ACP}%
\global\long\def\scoreFOne{F_{1}}%
\newcommandx\scoreFBeta[1][usedefault, addprefix=\global, 1=\beta]{F_{#1}}%
\global\long\def\scoreFOne{\scoreFBeta[1]}%
\global\long\def\priorpos{\pi_+}%
\global\long\def\priorneg{\pi_-}%
\global\long\def\scoreBiasIndex{BI}%
\global\long\def\ratepos{\tau_+}%
\global\long\def\rateneg{\tau_-}%
\global\long\def\scoreACP{ACP}%
\global\long\def\scorePFour{P_4}%
\global\long\def\normalizedConfusionMatrix{C}%
\global\long\def\scoreConfusionMatrixDeterminant{|\normalizedConfusionMatrix|}

\global\long\def\allEntities{\mathbb{E}}%
\global\long\def\entitiesToRank{\mathbb{E}}%
\global\long\def\anEntity{\epsilon}%
\global\long\def\randVarImportance{I}%
\global\long\def\randVarCanonicalImportance{\randVarImportance_{a,b}}
\global\long\def\canonicalRankingScore{\rankingScore[\randVarCanonicalImportance]}
\newcommandx\rankingScore[1][usedefault, addprefix=\global, 1=\randVarImportance]{R_{#1}}%
\global\long\def\canonicalRankingScore{\rankingScore[\randVarImportance_{a,b}]}
\global\long\def\scoreVUT{VUT}%
\global\long\def\tileCurvePriors{\gamma_\pi}
\global\long\def\tileCurveRates{\gamma_\tau}
\global\long\def\relWorseOrEquivalent{\lesssim}%
\global\long\def\relBetterOrEquivalent{\gtrsim}%
\global\long\def\relEquivalent{\sim}%
\global\long\def\relBetter{>}%
\global\long\def\relWorse{<}%
\global\long\def\relIncomparable{\not\lesseqqgtr}%
\global\long\def\rank{\mathrm{rank}_\entitiesToRank}%
\global\long\def\ordering{\relWorseOrEquivalent}%
\global\long\def\invertedOrdering{\relBetterOrEquivalent}%

\global\long\def\LScityscapes{\ding{171}}%
\global\long\def\LSade{\ding{170}}%
\global\long\def\LSvoc{\ding{169}}%
\global\long\def\LScoco{\ding{168}}%

\global\long\def\indicatorSymbol{\mathbf{1}}%
\global\long\def\realNumbers{\mathbb{R}}%
\global\long\def\aRelation{\mathcal{R}}%
\global\long\def\achievableByCombinations{\Phi}%
\global\long\def\allConvexCombinations{\mathrm{conv}}%
\newcommand{\indep}{\perp \!\!\! \perp}

\global\long\def\cityscapes{\LScityscapes{}~Cityscapes}
\global\long\def\ade{\LSade{}~ADE20K}
\global\long\def\voc{\LSvoc{}~Pascal VOC 2012}
\global\long\def\coco{\LScoco{}~COCO-Stuff 164k}

\newcommand{\MethodDesigner}{Bernadette}
\newcommand{\Benchmarker}{Leonard}
\newcommand{\AppDeveloper}{Howard}
\newcommand{\TheoreticalAnalyst}{Sheldon}

\newcommand{\tile}{Tile\xspace}
\newcommand{\tiles}{Tiles\xspace}
\newcommand{\valueTile}{Value Tile\xspace}
\newcommand{\valueTiles}{Value Tiles\xspace}
\newcommand{\baselineTile}{Baseline Value Tile\xspace}
\newcommand{\baselineTiles}{Baseline Value Tiles\xspace}
\newcommand{\SOTATile}{State-of-the-Art Value Tile\xspace}
\newcommand{\SOTATiles}{State-of-the-Art Value Tiles\xspace}
\newcommand{\noSkillTile}{No-Skill Tile\xspace}
\newcommand{\noSkillTiles}{No-Skill Tiles\xspace}
\newcommand{\skillTile}{Relative-Skill Tile\xspace}
\newcommand{\skillTiles}{Relative-Skill Tiles\xspace}
\newcommand{\correlationTile}{Correlation Tile\xspace}
\newcommand{\correlationTiles}{Correlation Tiles\xspace}
\newcommand{\rankingTile}{Ranking Tile\xspace}
\newcommand{\rankingTiles}{Ranking Tiles\xspace}
\newcommand{\entityTile}{Entity Tile\xspace}
\newcommand{\entityTiles}{Entity Tiles\xspace}

\global\long\def\aNonSkilledPerformance{\aPerformance_{\indep}}%
\global\long\def\allNonSkilledPerformances{\mathbb{\aPerformance}^{\randVarGroundtruthClass\indep\randVarPredictedClass}_{\measurableSpace}}%

\global\long\def\allPriorFixedPerformances{\mathbb{\aPerformance}^{\priorpos}_{\measurableSpace}}%

\newcommand{\comma}{\,,}
\newcommand{\point}{\,.}

\newcommandx\unconditionalProbabilisticScore[1]{\aScore_{#1}^{U}}%
\global\long\def\formulaPTN{\unconditionalProbabilisticScore{\eventTN}}%
\global\long\def\formulaPFP{\unconditionalProbabilisticScore{\eventFP}}%
\global\long\def\formulaPFN{\unconditionalProbabilisticScore{\eventFN}}%
\global\long\def\formulaPTP{\unconditionalProbabilisticScore{\eventTP}}%
\global\long\def\formulapriorneg{\unconditionalProbabilisticScore{\{\sampleTN,\sampleFP\}}}%
\global\long\def\formulapriorpos{\unconditionalProbabilisticScore{\{\sampleFN,\sampleTP\}}}%
\global\long\def\formularateneg{\unconditionalProbabilisticScore{\{\sampleTN,\sampleFN\}}}%
\global\long\def\formularatepos{\unconditionalProbabilisticScore{\{\sampleFP,\sampleTP\}}}%
\global\long\def\formulaAccuracy{\unconditionalProbabilisticScore{\{\sampleTN,\sampleTP\}}}%

\newcommandx\conditionalProbabilisticScore[2]{\aScore_{#1 \vert #2}^{C}}%
\global\long\def\formulaTNR{\conditionalProbabilisticScore{\{\sampleTN\}}{\{\sampleTN,\sampleFP\}}}%
\global\long\def\formulaTPR{\conditionalProbabilisticScore{\{\sampleTP\}}{\{\sampleFN,\sampleTP\}}}%
\global\long\def\formulaNPV{\conditionalProbabilisticScore{\{\sampleTN\}}{\{\sampleTN,\sampleFN\}}}%
\global\long\def\formulaPPV{\conditionalProbabilisticScore{\{\sampleTP\}}{\{\sampleFP,\sampleTP\}}}%
\global\long\def\formulaJaccardNeg{\conditionalProbabilisticScore{\{\sampleTN\}}{\{\sampleTN,\sampleFP,\sampleFN\}}}%
\global\long\def\formulaJaccardPos{\conditionalProbabilisticScore{\{\sampleTP\}}{\{\sampleFP,\sampleFN,\sampleTP\}}}%

\global\long\def\scoreBennettS{S_B}

\global\long\def\scoreSkewInsensitiveVersionFOne{SIVF}

\global\long\def\scoreSensitivity{Se}%
\global\long\def\scoreSpecificity{Sp}%
\global\long\def\scorePrecision{Pr}%
\global\long\def\scoreRecall{Re}%
\global\long\def\scoreIoU{IoU}%

\global\long\def\scorePrecisionPos{\scorePrecision^{+}}%
\global\long\def\scorePrecisionNeg{\scorePrecision^{-}}%
\global\long\def\scoreRecallPos{\scoreRecall^{+}}%
\global\long\def\scoreRecallNeg{\scoreRecall^{-}}%
\global\long\def\scoreIoUPos{\scoreIoU^{+}}%
\global\long\def\scoreIoUNeg{\scoreIoU^{-}}%
\global\long\def\scoreFBetaPos{\scoreFBeta^{+}}%
\global\long\def\scoreFBetaNeg{\scoreFBeta^{-}}%

\newcommand{\itn}{\randVarImportance_{\sampleTN}}
\newcommand{\ifp}{\randVarImportance_{\sampleFP}}
\newcommand{\ifn}{\randVarImportance_{\sampleFN}}
\newcommand{\itp}{\randVarImportance_{\sampleTP}}
\newcommand{\ptn}{\aPerformance_{\sampleTN}}
\newcommand{\pfp}{\aPerformance_{\sampleFP}}
\newcommand{\pfn}{\aPerformance_{\sampleFN}}
\newcommand{\ptp}{\aPerformance_{\sampleTP}}

\global\long\def\paramAllA{a}
\global\long\def\paramAllB{b}

\global\long\def\paramPriorsA{a^{(\pi)}}
\global\long\def\paramPriorsB{b^{(\pi)}}

\global\long\def\ourSpace{(\paramPriorsA,\paramPriorsB)}

\global\long\def\paramRatesA{a^{(\tau)}}
\global\long\def\paramRatesB{b^{(\tau)}}
\global\long\def\spearman{\rho}%
\global\long\def\kendall{\tau}%
\newcommand{\aDistance}{d}
\newcommand{\distSpearman}{\aDistance_\spearman}
\newcommand{\distKendall}{\aDistance_\kendall}

\newcommand{\optimality}{\mathcal{O}}

\global\long\def\meanA{\mu_A}%
\global\long\def\meanG{\mu_G}%
\global\long\def\meanH{\mu_H}%

\maketitle

\begin{abstract}
The accuracy, true negative rate (negative class recall), true positive rate (positive class recall), negative predictive value (negative class precision), positive predictive value (positive class precision), class-specific intersections over unions, class-specific F-scores, Cohen’s kappa, and balanced accuracy are scores that one intuitively wants to maximize when developing machine learning methods. However, they lead to different rankings of such methods. So, how can we establish a ranking that is an optimal tradeoff between any pair of these scores? It turns out that all these scores are special cases of ranking scores: a continuum of scores parameterized by the relative importance values assigned to the true negative, false positive, false negative, and true positive outcomes. Since all these scores are known to induce meaningful rankings, we seek the optimal tradeoffs among them. We consider the tradeoff between two scores to be optimal when it minimizes the Fréchet variance regarding the Kendall distance (also known as the bubble-sort distance) between rankings. For any given set of classifiers to rank, the optimal tradeoff lies at the center of the shortest (geodesic) path on the manifold of all rankings induced by the ranking scores. However, in the parameter space of the ranking scores, it is rarely achieved by averaging the importance values corresponding to the initial scores. In this paper, we present an exact analytical solution for determining a ranking score that is an optimal tradeoff for any given pair of initial scores, in the common case where the class priors are the same for all compared performances. We also experimentally investigate what happens when performance follows some distributions with fixed priors.
\end{abstract}

\section{Introduction}

\newcommand{\scoreA}{\rankingScore[\randVarImportance_A]}
\newcommand{\scoreB}{\rankingScore[\randVarImportance_B]}

Benchmarks aim at sorting some compared methods from best (first rank) to worst (largest rank). A common way to do this is to choose a score and consider that a method is \emph{worse than or equivalent to} another one if and only if the value taken by the score for the first method is \emph{less than or equal to} the value taken for the second method. In other words, using a score is nothing more than a trick to induce a preorder on the methods based on the well-known natural order on real numbers.

Obviously, not just any score or any preorder on performances can be used if the ranking is to be meaningful. When the performances have a probabilistic meaning, the conditions that must be satisfied have been formalized in \citet{Pierard2025Foundations}  (which presents \emph{generic axioms of performance-based ranking}) without assuming any particular task. Moreover, in the ubiquitous setting in which the evaluation behaves linearly regarding the evaluated method (\ie, a hybrid method obtained by blindly and non-deterministically combining some base methods has a performance that is a mixture of the performances of the base methods), it has been shown that a wide family of scores leads to meaningful rankings. These are called \emph{ranking scores} and defined as the ratio of the expected value of the importance-satisfaction product to the expected value of the importance, the \emph{satisfaction} being a random variable modeling the task and the \emph{importance} being another random variable specifying the relative importance given to the various evaluation outcomes \citep{Pierard2025Foundations}.

In other words, the ranking scores $\rankingScore$ form an infinite and continuous family of scores parameterized by importance values. This offers great flexibility in freely choosing the relative weights that best align with the preferences associated with the target application or its users. However, we argue that choosing the score (\ie, the importance values in the case of ranking scores) remains a delicate matter for a challenge organizer or a benchmarker \citep{Halin2024AHitchhikers-arxiv}, for example, who is unaware of the applications that the people consulting the proposed ranking are targeting. Consequently, further theoretical advances to guide this choice are still needed. This paper aims to contribute to that effort for two-class classification, with a special focus on the particular case in which all compared performances share the same class priors. This is a very common setting as good practice suggests to compare only performances determined by evaluating the methods on the same dataset, hence fixing the class priors.

In two-class classification, one usually considers a class as \emph{negative} and the other one as \emph{positive}. It should be stressed that this is just a convention. The choice of which class is positive is always arbitrary. Four evaluation outcomes are usually considered: true negative ($\sampleTN$), false positive ($\sampleFP$), false negative ($\sampleFN$), and true positive ($\sampleTP$). All of us will certainly agree that the outcomes $\sampleTN$ and $\sampleTP$ are fully satisfying, and that the outcomes $\sampleFP$ and $\sampleFN$ are not satisfying at all. In this setting, the ranking scores include well-known scores such as the \emphScore{accuracy} $\scoreAccuracy$, the \emphScore{true negative rate} $\scoreTNR=\scoreRecallNeg=\scoreSpecificity$ (\aka\ \emphScore{inverse recall} and \emphScore{specificity}), \emphScore{the true positive rate} $\scoreTPR=\scoreRecallPos=\scoreSensitivity$ (\aka\ \emphScore{recall} and \emphScore{sensitivity}), the \emphScore{negative predictive value} $\scoreNPV=\scorePrecisionNeg$ (\aka\ \emphScore{inverse precision}), the \emphScore{positive predictive value} $\scorePPV=\scorePrecisionPos$ (\aka\ \emphScore{precision}), as well as the \emphScore{F-scores} $\scoreFBeta$ and the \emphScore{intersection over union} $\scoreIoU$ (\aka\ \emphScore{Jaccard coefficient}) that can be defined for both classes. When the class priors are fixed, the \emphScore{balanced accuracy} $\scoreBalancedAccuracy$, the \emphScore{skew-insensitive version of $\scoreFOne$} $\scoreSkewInsensitiveVersionFOne$~\citep{Flach2003TheGeometry}, and the \emphScore{standardized predictive values} $\scoreSNPV$ and $\scoreSPPV$~\citep{Heston2011Standardizing}  also belong to the family of ranking scores (the importance values are constants expressed in terms of the class priors). Moreover, \emphScore{Cohen's kappa} $\scoreCohenKappa$ \citep{Cohen1960ACoefficient} (\aka\ \emphScore{Heidke Skill Score}) is linearly increasing with a ranking score \citep{Pierard2024TheTile-arxiv}. All these classical scores are therefore associated with well-defined importance values. For any ranking score, its inverse (\ie, the function composition of the score with the operation on performances that swaps the negative and positive classes~\citep{Powers2020Evaluation-arxiv}) is also a ranking score~\citep{Pierard2024TheTile-arxiv}. For example, the true negative and positive rates are inverses of each other, and the negative and positive predictive values are also inverses of each other.

The main objective of this paper is to provide all the theoretical and algorithmic tools to determine optimal tradeoffs between two scores. This indeed includes the determination of the tradeoff between a class-specific score (\eg, the precision, recall, F-score, or intersection over union) and its inverse. But this also includes the determination of the tradeoff between two scores for the same class (\eg between precision and recall).

There are mainly two reasons why one cannot just use any arbitrarily chosen mean (\eg, arithmetic, geometric, harmonic, or quadratic) to establish a tradeoff between scores. First, the performance ordering induced by a mean of two base scores does not necessarily satisfy the axioms of performance-based ranking, even if the orderings induced by these two base scores do. Second, even if this were the case, the resulting ranking would not necessarily be intermediate between the rankings induced by the two base scores. For example, it has been shown in \citet{Pierard2026What} that even though the F-score $\scoreFOne$ is the harmonic mean of the precision and recall and that $\scoreFOne$ is known to lead to meaningful rankings (it is a ranking score), the F-score sometimes mimics perfectly the precision or recall in terms of rankings, totally ignoring the other. It is very likely that this problem is not specific to the tradeoff between precision and recall. For this reason, it is far preferable to establish an optimal tradeoff from the perspective of ranking. We aim at providing all the theoretical and algorithmic tools to answer the following question.
\begin{question*}
    In two-class classification, for any given set of performances and for any pair of scores perfectly rank-correlated with ranking scores (\eg, $\scoreAccuracy$, $\scoreTNR$, $\scoreTPR$, $\scoreNPV$, $\scorePPV$, $\scoreFBetaNeg$, $\scoreFBetaPos$, $\scoreIoUNeg$, $\scoreIoUPos$, $\scoreBalancedAccuracy$, %
    $\scoreSkewInsensitiveVersionFOne$, $\scoreCohenKappa$), what ranking scores are optimal tradeoffs from a ranking point of view?
\end{question*}
To the best of our knowledge, only the very particular case of the optimal tradeoff between the precision and recall of the positive class has been extensively studied. This was done in \citet{Pierard2026What}. 
In this paper, we keep the same methodology and reasoning, generalizing it to any pair of ranking scores. Let us summarize the methodology in four points.

\textbf{(1) The search space.}
How one interprets a ranking depends on how one interprets the score that induces it. For this reason, the use of rank aggregation methods, such as those developed by Kemeny~\citep{Kemeny1959Mathematics}, are inappropriate as they operate on the rankings only, ignoring their origins, and lead to a ranking that is not related to any score. Instead, we explicitly determine the tradeoff score that leads to the optimal tradeoff ranking. Moreover, as we want to be sure that the tradeoff corresponds to one of the possible meaningful rankings, we seek for the optimal tradeoff within the family of ranking scores. More information about the search space can be found in \cref{sec:search-space}.

\textbf{(2) The means.}
The main challenge facing us is that the reasoning of \citet{Pierard2026What} was heavily based on the observation that there exists a weighted mean (the weighted harmonic mean) such that, if applied to precision and recall, one obtains ranking scores. To be able to generalize it, for any pair of ranking scores we need to find a weighted mean such that, if applied to them, one also obtains a ranking score. This is done in \cref{sec:means}.

\textbf{(3) The shortest paths.}
If there is a mean between the scores $\scoreA$ and $\scoreB$, then it is possible to determine scores $\rankingScore$ such that $\distKendall(\scoreA;\rankingScore) + \distKendall(\rankingScore;\scoreB) = \distKendall(\scoreA;\scoreB)$ where $\distKendall=\frac{1-\kendall}{2}$ is Kendall distance \citep{Kendall1938ANewMeasure} (also known as the bubble-sort distance) between rankings. We say that these scores form a shortest path between $\scoreA$ and $\scoreB$. This is done in \cref{sec:shortest-paths}.

\textbf{(4) The optimal tradeoffs.}
The optimal tradeoff is the midpoint on the shortest path. It is given by the ranking score $\rankingScore$ such that $\distKendall(\scoreA;\rankingScore) = \distKendall(\rankingScore;\scoreB)$. It is a Karcher mean \citep{Karcher1977Riemannian} as it minimizes the Fréchet variance regarding the Kendall distance between rankings:
\begin{equation}
    \rankingScore[\randVarImportance^*]
    = \underset{\rankingScore}{\arg\min} \, \sigma^2(\randVarImportance)
    \qquad
    \textrm{with}
    \qquad
    \sigma^2(\randVarImportance) = \distKendall^2(\scoreA;\rankingScore) + \distKendall^2(\rankingScore ; \scoreB)
    \point
    \label{eq:Karcher-mean}
\end{equation}
Establishing solutions for this equation is done in \cref{sec:optimal-tradeoffs}.

Our theoretical contributions are threefold.
\begin{enumerate}
    \item From the point of view of \emph{values}, we introduce scores that are homographic transformations of ranking scores. We show that, when the class priors are fixed, for any pair of ranking scores $(\scoreA,\scoreB)$ there exist ranking scores that are weighted $f$-means of some ranking scores $\scoreA'$ and $\scoreB'$ monotonically increasing with respect to $\scoreA$ and $\scoreB$, respectively, and that are generated by homographies $f$. This stands in contrast to the unjustifiable practice of considering only the arithmetic, geometric, and harmonic means.
    \item From the point of view of \emph{rankings}, we provide the equation for a shortest path between any pair of ranking scores $(\scoreA,\scoreB)$, assuming the priors are fixed. Furthermore, we present a method for determining the optimal tradeoff when the set of performances to be ranked is known. This optimal tradeoff lies on the shortest path.
    \item We show that, among all the ways to parameterize the two-dimensional manifold of performance orderings induced by ranking scores, there is one that has properties particularly interesting for finding optimal tradeoffs. We call this \emph{the parameterization adapted to the priors}. Its usefulness should not be viewed as limited to simplifying equations, since a well-designed parameterization is key to obtaining ranking visualization tools that are well-suited to specific needs. Readers curious to discover such a tool, called \emph{the Tile}, are invited to consult \citet{Pierard2024TheTile-arxiv,Pierard2025Multidomain-arxiv} for the theory, \citet{Pierard2025Sorbetto-zenodo} for an open-source implementation, and \citet{Halin2024AHitchhikers-arxiv,Pierard2025AMethodology,Bouwmans2026Illegal,Zakharova2026Event-arxiv} for some uses of it.
\end{enumerate}

\section{Related Work: The Risky Means}

We dedicate this section to the tradeoffs between \emph{two} scores that have been defined in the literature, in the case of two-class classification. As this list shows, these tradeoffs are all established using an arbitrarily chosen mean. The so-called “Pythagorean” means (\ie, the arithmetic mean $\meanA$, the geometric mean $\meanG$, and the harmonic mean $\meanH$) are systematically used, although there is no good reason to limit ourselves to the means that were known and studied by the ancient Greeks, some two and a half millennia ago. The base scores on which these tradeoffs are based are primarily the following: $\scoreTNR$, $\scoreTPR$, $\scoreNPV$, and $\scorePPV$. Hereafter, we list not only the means but also their monotonous transformations as they induce the same ranking or its inverse.
\emphTradeoff{Scores related to $\meanA(\scoreTNR;\scoreTPR)$.} 
Many authors combine $\scoreTNR$ and $\scoreTPR$. Their weighted arithmetic mean is the \emphScore{weighted accuracy} $\scoreWeightedAccuracy$. When the weights are $0.5$, we obtain the \emphScore{balanced accuracy} $\scoreBalancedAccuracy$ \citep{Brodersen2010TheBalanced}, and when the weights correspond to the priors, we get the \emphScore{accuracy} $\scoreAccuracy$. \emphScore{Youden's index} \citep{Youden1950Index,Fluss2005Estimation} or \emphScore{Youden's} $\scoreYoudenJ$ \emphScore{statistic} is $\scoreYoudenJ=2\scoreBalancedAccuracy-1$ . It is also called \emphScore{informedness} \citep{Powers2020Evaluation-arxiv} and \emphScore{Peirce Skill Score}~\citep{Peirce1884TheNumerical}. The \emphScore{determinant of the (normalized) confusion matrix}~\citep{Foulle2025Mathematical} is $\scoreConfusionMatrixDeterminant=\priorneg\priorpos\scoreYoudenJ$.
\emphTradeoff{Scores related to $\meanG(\scoreTNR;\scoreTPR)$.}
Instead of taking an arithmetic mean, it has been proposed to take the geometric mean~\citep{Kubat1997Addressing,Barandela2003Strategies,Guo2008OnTheClass}, which leads to the \emphScore{G-measure}~\citep{Canbek2017Binary}. %
\emphTradeoff{Scores related to $\meanH(\scorePPV;\scoreTPR)$.}
This is the \emphScore{F-one} score \citep{VanRijsbergen1979Information,Christen2023AReview}, also called \emphScore{Dice-S{\o}rensen coefficient} \citep{Dice1945Measures}. It is well known that that Jaccard coefficient \citep{Jaccard1908Nouvelles} leads to the same performance ordering as $\scoreFOne$, due to a strict increasing monotonic relationship between the two scores.
\emphTradeoff{$\meanG(\scorePPV;\scoreTPR)$} is \emphScore{Fowlkes-Mallows index} \citep{Fowlkes1983AMethod}.
\emphTradeoff{$\meanA(\scorePPV;\scoreTPR)$} is \emphScore{Kulczynski index} \citep{Kulczynski1927DiePflanzenassoziationen}. The \emphScore{Individual Classification Success Index (ICSI)} is a monotonous transformation of Kulczynski index.
\emphTradeoff{Scores related to $\meanA(\scoreNPV;\scorePPV)$.}
The \emphScore{markedness}~\citep{Powers2020Evaluation-arxiv} is defined as $\scoreNPV+\scorePPV-1$ and is also known as the \emphScore{Clayton Skill Score}~\citep{Clayton1934Rating,Canbek2017Binary,Wilks2020Statistical}.

\section{Theory}

\paragraph{Probabilistic grounding and notations (following \citet{Pierard2025Foundations}).}
Let us consider a measurable space $\measurableSpace$ with a sample space (\aka universe) made of four evaluation outcomes $\sampleSpace=\{\sampleTN,\sampleFP,\sampleFN,\sampleTP\}$ and %
the event space $\eventSpace=2^\sampleSpace$. By definition, performances are probability measures on $\sampleSpace$, %
and scores are real functions whose domains are subsets of all possible performances. We denote by $\expectedValueSymbol_\aPerformance[\cdot]$ the expected value of random variables for the performance $\aPerformance$. In order to establish a connection with two-class classification, we introduce a set $\allClasses=\{\classNeg,\classPos\}$ that we interpret as the set of classes, $\classNeg$ being called the \emph{negative class} and $\classPos$ being called the \emph{positive class}. Moreover, since it is common to refer to ground-truth and predicted classes, we introduce symbolic variables for them. The ground-truth class is $\randVarGroundtruthClass:\sampleSpace\rightarrow\allClasses$ with $\randVarGroundtruthClass(\sampleTN)=\randVarGroundtruthClass(\sampleFP)=\classNeg$ and $\randVarGroundtruthClass(\sampleFN)=\randVarGroundtruthClass(\sampleTP)=\classPos$. The predicted class is $\randVarPredictedClass:\sampleSpace\rightarrow\allClasses$ with $\randVarPredictedClass(\sampleTN)=\randVarPredictedClass(\sampleFN)=\classNeg$ and $\randVarPredictedClass(\sampleFP)=\randVarPredictedClass(\sampleTP)=\classPos$. The symbols $\sampleTN$, $\sampleFP$, $\sampleFN$, and $\sampleTP$ can thus be interpreted, respectively, as a \emph{true negative}, \emph{false positive}, \emph{false negative}, and \emph{true positive}. For a performance $\aPerformance$, the class priors are given by $\priorneg=\aPerformance(\randVarGroundtruthClass=\classNeg)$ and $\priorpos=\aPerformance(\randVarGroundtruthClass=\classPos)$. %
The classification task is modeled by a binary random variable called \emph{satisfaction} $\randVarSatisfaction:\sampleSpace\rightarrow\realNumbers:\aSample\mapsto\indicatorSymbol_{\{\sampleTN,\sampleTP\}}(\aSample)$. In other words $\randVarSatisfaction=1$ if and only if $\randVarGroundtruthClass=\randVarPredictedClass$ and $\randVarSatisfaction=0$ if and only if $\randVarGroundtruthClass\ne\randVarPredictedClass$. We also introduce another random variable, positive but non-zero, called \emph{importance} $\randVarImportance:\sampleSpace\rightarrow[0,\infty[$. To shorten the notations, we pose $\randVarImportance_{\aSample}=\randVarImportance(\aSample)$ and $\aPerformance_{\aSample}=\aPerformance(\{\aSample\})$ for all $\aSample\in\sampleSpace$.

\subsection{The Performance Orderings Induced by Ranking Scores}
\label{sec:search-space}

\begin{table}
    \tiny
    \centering
    \begin{tabular}{ll|cccc|cc}
        score & 
        classic formula & 
        $\itn$ & $\ifp$ & $\ifn$ & $\itp$ &
        $\paramPriorsA(\randVarImportance)$ & $\paramPriorsB(\randVarImportance)$\\
        \hline
        $\scoreTNR=\scoreRecallNeg$ & 
        $\aPerformance(\randVarGroundtruthClass=\randVarPredictedClass|\randVarGroundtruthClass=\classNeg)$ & 
        $1$ & $1$ & $0$ & $0$ &
        $0$ & $0$ \\
        $\scoreTPR=\scoreRecallPos$ & 
        $\aPerformance(\randVarGroundtruthClass=\randVarPredictedClass|\randVarGroundtruthClass=\classPos)$ & 
        $0$ & $0$ & $1$ & $1$ &
        $1$ & $1$ \\
        $\scoreNPV=\scorePrecisionNeg$ & 
        $\aPerformance(\randVarGroundtruthClass=\randVarPredictedClass|\randVarPredictedClass=\classNeg)$ & 
        $1$ & $0$ & $1$ & $0$ &
        $0$ & $1$ \\
        $\scorePPV=\scorePrecisionPos$ & 
        $\aPerformance(\randVarGroundtruthClass=\randVarPredictedClass|\randVarPredictedClass=\classPos)$ & 
        $0$ & $1$ & $0$ & $1$ &
        $1$ & $0$ \\
        $\scoreSNPV$ & 
        $\frac{ TNR }{ TNR + FNR }=\frac{ NPV \pi_+ }{ NPV ( \pi_+ - \pi_- ) + \pi_- }$ & 
        $\priorpos$ & $0$ & $\priorneg$ & $0$ &
        $0$ & $1$ \\
        $\scoreSPPV$ & 
        $\frac{ TPR }{ FPR + TPR }=\frac{ PPV  \pi_- }{ PPV ( \pi_- - \pi_+ ) + \pi_+ }$ & 
        $0$ & $\priorpos$ & $0$ & $\priorneg$ &
        $1$ & $0$ \\
        $\scoreSkewInsensitiveVersionFOne$ & 
        $\frac{2 \, \scoreTPR}{\scoreTPR + \scoreFPR + 1} = \mu_H(\scoreSPPV;\scoreTPR)$ & 
        $0$ & $\priorpos$ & $\priorneg$ & $2 \, \priorneg$ &
        $1$ & $\nicefrac12$ \\
        $\scoreFBetaNeg$ & 
        $\frac{1}{\scoreFBetaNeg} = \frac{1}{1+\beta^2} \frac{1}{\scorePrecisionNeg} + \frac{\beta^2}{1+\beta^2} \frac{1}{\scoreRecallNeg}$ & 
        $1$ & $\frac{\beta^2}{1+\beta^2}$ & $\frac{1}{1+\beta^2}$ & $0$ &
        $0$ & $\frac{1}{1+\ell}$, $\ell=\beta^2\frac{\priorneg}{\priorpos}$\\
        $\scoreFBetaPos$ & 
        $\frac{1}{\scoreFBetaPos} = \frac{1}{1+\beta^2} \frac{1}{\scorePrecisionPos} + \frac{\beta^2}{1+\beta^2} \frac{1}{\scoreRecallPos}$ & 
        $0$ & $\frac{1}{1+\beta^2}$ & $\frac{\beta^2}{1+\beta^2}$ & $1$ &
        $1$ & $\frac{\ell}{1+\ell}$, $\ell=\beta^2\frac{\priorpos}{\priorneg}$\\
        $\scoreIoUNeg$ & 
        $\aPerformance(\randVarGroundtruthClass=\randVarPredictedClass|\randVarGroundtruthClass=\classNeg \cup \randVarPredictedClass=\classNeg)$ & 
        $1$ & $1$ & $1$ & $0$ &
        $0$ & $\priorpos$ \\
        $\scoreIoUPos$ & 
        $\aPerformance(\randVarGroundtruthClass=\randVarPredictedClass|\randVarGroundtruthClass=\classPos \cup \randVarPredictedClass=\classPos)$ & 
        $0$ & $1$ & $1$ & $1$ &
        $1$ & $\priorpos$ \\
        $\scoreBalancedAccuracy$ & 
        $\nicefrac12 \, \scoreTNR + \nicefrac12 \, \scoreTPR$ &
        $\priorpos$ & $\priorpos$ & $\priorneg$ & $\priorneg$ &
        $\nicefrac12$ & $\nicefrac12$ \\
        $\scoreAccuracy$ & 
        $\aPerformance(\randVarGroundtruthClass=\randVarPredictedClass)=\expectedValueSymbol_\aPerformance[\randVarSatisfaction]$ & 
        $1$ & $1$ & $1$ & $1$ &
        $\priorpos$ & $\priorpos$ \\
        Linear of $\scoreCohenKappa$ &
        $(\priorneg^{2}+\priorpos^{2})\scoreCohenKappa+2\priorneg\priorpos$ & 
        $\frac{\priorpos^{2}}{\priorneg^{2}+\priorpos^{2}}$ & $\frac{1}{2}$ & $\frac{1}{2}$ & $\frac{\priorneg^{2}}{\priorneg^{2}+\priorpos^{2}}$ &
        $\priorneg$ & $\priorpos$ \\
    \end{tabular}
    \caption{Some ranking scores, when the class priors $(\priorneg,\priorpos)$ are fixed, with corresponding importance values, and the coordinates of the induced performance ordering in the $(\paramPriorsA,\paramPriorsB)$ space.}
    \label{tbl:ranking-scores}
\end{table}

\begin{figure}
    \centering
    \includegraphics[width=\linewidth]{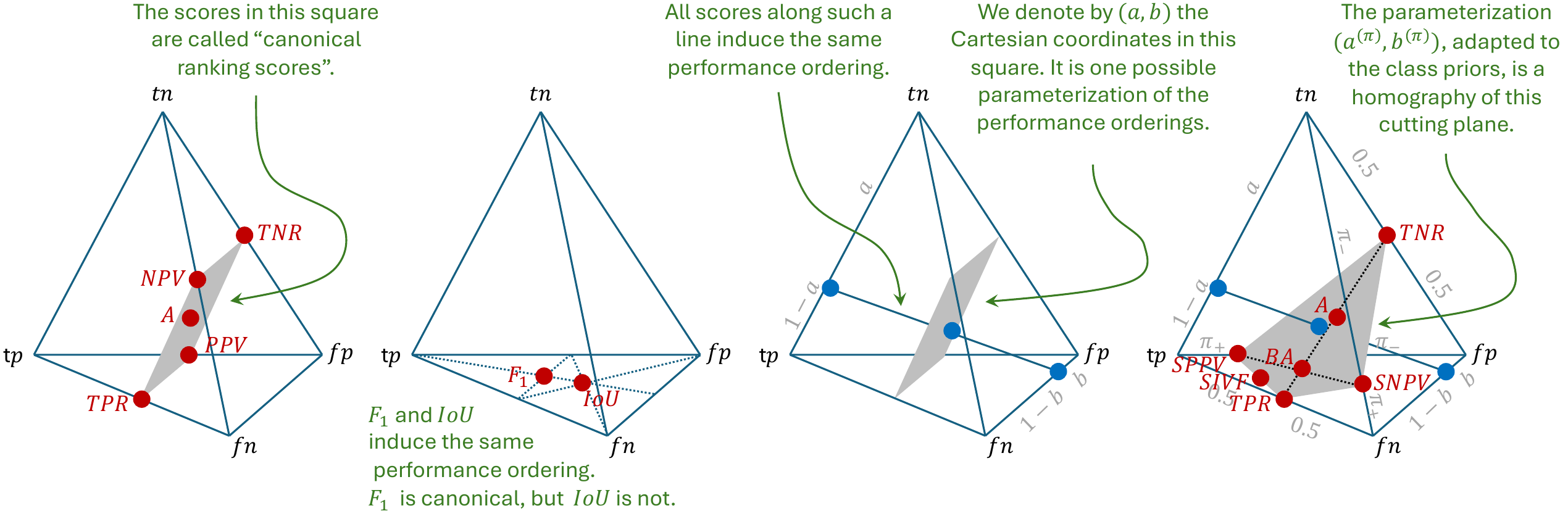}
    \caption{The tetrahedron of ranking scores, or the simplex of importance values in two-class classification. Be careful not to confuse this with the tetrahedron of performances or confusion simplex~\citep{Chicco2021TheBenefits,Chicco2023TheMatthews,Lovell2023NeverMind,Brzezinski2017Tetrahedron,Brzezinski2020OnTheDynamics,Brzezinski2018Visualbased}. In this one, each point corresponds to a score, not a performance.}
    \label{fig:tetrahedron-ranking-scores}
\end{figure}

\paragraph{A space of scores.}
To ensure meaningful ranking tradeoffs, we seek for optimal tradeoffs within the family of ranking scores. These scores are generic, but we are only interested in their particularization to two-class classification. They have been defined in \citet{Pierard2025Foundations} as
\begin{equation}
    \rankingScore(\aPerformance)
    = \frac{
        \expectedValueSymbol_\aPerformance[\randVarSatisfaction\randVarImportance]
    }{
        \expectedValueSymbol_\aPerformance[\randVarImportance]
    }
    = \frac{
        \itn \, \ptn + \itp \, \ptp
    }{
        \itn \, \ptn + \ifp \, \pfp + \ifn \, \pfn + \itp \, \ptp
    }
    \point
\end{equation}
A few examples of ranking scores or of scores strictly increasing with them (\ie, perfectly rank-correlated with them) are given in \cref{tbl:ranking-scores}.
The ranking scores are insensitive to the scaling of the importance $\randVarImportance$, so we can limit ourselves to considering that $\itn+\ifp+\ifn+\itp=1$. As the importance values must be positive, the ranking scores form a three-dimensional simplex (see \cref{fig:tetrahedron-ranking-scores}). In the tetrahedron $(\itn,\ifp,\ifn,\itp)$, all scores on the line between $(1-\paramAllA,0,0,\paramAllA)$ and $(0,1-\paramAllB,\paramAllB,0)$ induce the same performance ordering. We call them \emph{iso-ranking lines}.

\paragraph{A space of performance orderings.}
The performance orderings can be parameterized in different ways. A quite intuitive choice is with the pair $(\paramAllA, \paramAllB)\in[0,1]^2$. These are Cartesian coordinates in the square $\itn+\itp=\ifp+\ifn$ cutting the tetrahedron in the middle and intersecting each iso-ranking line in one point. But we discovered that there is a much more interesting cutting plane to consider for our purpose: $\itn \, \priorneg + \itp \, \priorpos = \ifp \, \priorneg + \ifn \, \priorpos$. Its advantages will become clear in the next subsections. This plane form a convex kite %
that can be mapped to the square $[0,1]^2$ with a homography %
(thus a collineation; this will be important for us a little bit further) %
$\ourSpace$:
\begin{align}
    \paramPriorsA(\randVarImportance) &= \frac{
        \itp \, \priorpos
    }{
        \itn \, \priorneg + \itp \, \priorpos
    } \in[0,1]
    & \paramPriorsB(\randVarImportance) &= \frac{
        \ifn \, \priorpos
    }{
        \ifp \, \priorneg + \ifn \, \priorpos
    } \in[0,1] \point
    \label{eq:paramPriors}
\end{align}
Note that $\paramPriorsA$ and $\paramPriorsB$ are strictly increasing with $\paramAllA$ and $\paramAllB$, respectively.

\subsection{Means of Ranking Scores}
\label{sec:means}

Let us remind that for real numbers (the co-domain of scores), a large class of means is known as the generalized $f$-means, quasi-arithmetic means, or Kolmogorov-Nagumo-de Finetti means~\citep{Kolmogorov1930SurLaNotionDeLaMoyenne}. The generalized $f$-mean $\overline{\aScore}$, on an interval $[l,u]$, between $\aScore_1$ and $\aScore_2$ is defined as $f(\overline{\aScore})=\frac{1}{2} f(\aScore_1) + \frac{1}{2} f(\aScore_2)$, where $f$ is monotonic over $[l,u]$. These means include the arithmetic, geometric, harmonic, and quadratic means, to cite only a few examples. In the case of a binary satisfaction, in which we are, it has been shown in \citet{Pierard2025Foundations} that, for the generators $f:x\mapsto x^{-1}$ and $f:x\mapsto (1-x)^{-1}$, some ranking scores are means of others. When the priors are fixed, we discovered that the means to consider are more generally the $f$-means generated by some homographies $f$.\footnote{If two weighted means are equal, their generators are related by homographies \citep{Grunwald2019OnTheEquality}. Also, the composition of two homographies is also a homography. That is why we consider all homographies.} For this reason, we introduce the homographic transformations %
of ranking scores:
\begin{equation}
    f \circ \rankingScore
    = \frac{
        A \, \rankingScore + B
    }{
        C \, \rankingScore + D
    }
    = \frac{
        (A+B) \, \itn \, \ptn 
        + B \, \ifp \, \pfp 
        + B \, \ifn \, \pfn 
        + (A+B) \, \itp \, \ptp
    }{
        (C+D) \, \itn \, \ptn 
        + D \, \ifp \, \pfp 
        + D \, \ifn \, \pfn 
        + (C+D) \, \itp \, \ptp
    }
    \comma
    \label{eq:homographic-transfo-ranking-score}
\end{equation}
with $A D > B C$. This constraint on the transformation parameters ensures that the transformation is strictly increasing and therefore that the score $f \circ \rankingScore$ induces the same performance ordering as $\rankingScore$.

\newcommand{\ditn}{\Delta_{\sampleTN}}
\newcommand{\difp}{\Delta_{\sampleFP}}
\newcommand{\difn}{\Delta_{\sampleFN}}
\newcommand{\ditp}{\Delta_{\sampleTP}}

\newcommand{\da}{\Delta_{\paramPriorsA}}
\newcommand{\db}{\Delta_{\paramPriorsB}}

From now, we leave aside the unnecessary constraint $\sum_\aSample \randVarImportance(\aSample)=1$. To ensure $\itn \, \priorneg + \itp \, \priorpos = \ifp \, \priorneg + \ifn \, \priorpos$, for the performance ordering $(\paramPriorsA,\paramPriorsB)$, we consider only
\begin{align}
    \itn &= \priorpos \, (1-\paramPriorsA) &
    \ifp &= \priorpos \, (1-\paramPriorsB) &
    \ifn &= \priorneg \, \paramPriorsB &
    \itp &= \priorneg \, \paramPriorsA
    \point
    \label{eq:our-plane}
\end{align}

We now take a look at the conditions on $(A,B,C,D)$ to obtain a score $\rankingScore$ with the $f$-mean of some given scores $\scoreA$ and $\scoreB$. The equation $f \circ \rankingScore = \nicefrac12 f \circ \scoreA + \nicefrac12 f \circ \scoreB$ has solutions when $\randVarImportance$ is the arithmetic mean of $\randVarImportance_A$ and $\randVarImportance_B$ and the denominator of \cref{eq:homographic-transfo-ranking-score} is the same for $f \circ \scoreA$ and $f \circ \scoreB$, for any given performance. Thus,
$%
    (C+D) \, \ditn \, \ptn 
    + D \, \difp \, \pfp 
    + D \, \difn \, \pfn 
    + (C+D) \, \ditp \, \ptp = 0
$, %
with $\Delta_\aSample=\randVarImportance_A(\aSample)-\randVarImportance_B(\aSample) \, \forall\aSample\in\sampleSpace$. As $\ptn+\pfp$ and $\pfn+\ptp$ are constant when the priors are fixed, the coefficients in front of $\ptn$ and $\pfp$ should be equal, as well as those of $\pfn$ and $\ptp$:
\begin{align}
    (C+D) \, \ditn &= D \, \difp &
    D \, \difn &= (C+D) \, \ditp
    \label{eq:twoConstraints}
\end{align}
Let us pose $\da=\paramPriorsA(\randVarImportance_B)-\paramPriorsA(\randVarImportance_A)$ and $\db=\paramPriorsB(\randVarImportance_B)-\paramPriorsB(\randVarImportance_A)$. We have 
$\ditn = - \priorpos \, \da$, 
$\difp = - \priorpos \, \db$, 
$\difn = \priorneg \, \db$, 
and $\ditp = \priorneg \, \da$. So, the two constraints given in \cref{eq:twoConstraints} are equivalent to $(C+D) \, \da = D \, \db$. We choose the solution
\begin{align}
    A & = \da &
    B &= \da - \db &
    C &= \db - \da &
    D &= \da
    \point
    \label{eq:homography-coefficients}
\end{align}
This homography $f$ is an involution (\ie, $f^{-1}=f$). %
Its pole $z=-\nicefrac{D}{C}$ can be in the $[0,1]$ range (\ie, the image of ranking scores). So, $f$ is monotonously increasing on $]-\infty,z[$ and on $]z,\infty[$, but not on $[0,1]$. However, this is not a problem for us because it can be shown that
\begin{align}
    \scoreA(\aPerformance) < z &\Leftrightarrow \scoreB(\aPerformance) < z &
    \scoreA(\aPerformance) = z &\Leftrightarrow \scoreB(\aPerformance) = z &
    \scoreA(\aPerformance) > z &\Leftrightarrow \scoreB(\aPerformance) > z
    \point
\end{align}

\subsection{Universal Shortest Paths Between Rankings}
\label{sec:shortest-paths}

We have just seen that, when the class priors are fixed, for any pair of ranking scores $(\scoreA,\scoreB)$ there exist ranking scores $\rankingScore$ that are weighted means of some ranking scores $\scoreA'$ and $\scoreB'$ monotonically increasing with respect to $\scoreA$ and $\scoreB$, respectively. The reasoning of \citet{Pierard2026What} can be applied as is: it follows immediately that $\distKendall(\scoreA;\rankingScore) + \distKendall(\rankingScore;\scoreB) = \distKendall(\scoreA;\scoreB)$ for all these means $\rankingScore$. This shortest path is obtained with a weighted linear interpolation of importance values when they satisfy \cref{eq:our-plane}. However, taking $\randVarImportance = \frac12 \randVarImportance_A + \frac12 \randVarImportance_B$ does not necessarily lead to optimal tradeoff (this will be the topic of the next subsection). It is interesting to note that the passage from the plane of \cref{eq:our-plane} to the cutting plane shown on the right hand side tetrahedron of \cref{fig:tetrahedron-ranking-scores} is a homography. As the mapping from this cutting plane to the space $\ourSpace$ is also a homography, it follows that the shortest paths are line segments in $\ourSpace$. Aside from the ease of graphical representation this provides (see \cref{sec:experiments}), it also sheds light on the manifolds of performance orderings induced by the ranking scores when the class priors are fixed. According to Beltrami's theorem \citep{Beltrami1868Teoria}, this implies that these manifolds have a constant curvature. %

\subsection{Optimal Tradeoffs Between Two Ranking Scores}
\label{sec:optimal-tradeoffs}

Again, the reasoning that was proposed in \cite{Pierard2026What} to determine the optimal tradeoff between $\scorePrecision$ and $\scoreRecall$ can be easily generalized to the determine the optimal tradeoff between any pair of ranking scores thanks to the fact that we found means. %
If there is a score $\rankingScore$ such that $\distKendall(\scoreA;\rankingScore) = \distKendall(\rankingScore;\scoreB)$, then it is a solution to \cref{eq:Karcher-mean}.

Let us consider any score $\aScore$ that is a weighted $f$-mean of some scores $\aScore_1$ and $\aScore_2$, that is $f \circ \aScore = (1-w) \, f \circ \aScore_1 + w \, f \circ \aScore_2$ with $w \in [0,1]$. The performances $\aPerformance_i$ and $\aPerformance_j$ are put on an equal footing by $\aScore$ when
\begin{equation}
    \aScore(\aPerformance_i) = \aScore(\aPerformance_j)
    \Leftrightarrow
    \frac{w}{w-1}
    = \frac{
        ( f \circ \aScore_1 ) ( \aPerformance_i ) - ( f \circ \aScore_1 ) ( \aPerformance_j )
    }{
        ( f \circ \aScore_2 ) ( \aPerformance_i ) - ( f \circ \aScore_2 ) ( \aPerformance_j )
    }
    = \psi ( \aPerformance_i; \aPerformance_j )
    \point
    \label{eq:weight-for-boundry}
\end{equation}
The weight $w$ depends on the pair of performances $( \aPerformance_i; \aPerformance_j )$. We have $w ( \aPerformance_i; \aPerformance_j ) =\frac{\psi ( \aPerformance_i; \aPerformance_j )}{\psi ( \aPerformance_i; \aPerformance_j ) - 1}$. If both $\aScore_1$ and $\aScore_2$ agree on the order between $\aPerformance_i$ and $\aPerformance_j$, then $w ( \aPerformance_i; \aPerformance_j ) \not\in [0,1]$.

The optimal tradeoffs between $\aScore_1$ and $\aScore_2$, from a ranking perspective, are the scores on the shortest path such that there is the same amount of swaps on both sides. Denoting multisets by $\{\cdot\}$, a weight leading to an optimal tradeoff is given by
\begin{equation}
    w^* = \mathrm{median} \left( \left\{ w(\aPerformance_i,\aPerformance_j) \, \vert \, i \ne j \wedge w(\aPerformance_i,\aPerformance_j) \in [0,1] \right\} \cup \left\{0,1\right\} \right)
    \point
    \label{eq:closed-form-expression}
\end{equation}

\section{Algorithm}

We present now an algorithm in nine steps to determine a ranking score that is one of the optimal tradeoffs between two base scores, namely $\aScore_A$ and $\aScore_B$, perfectly rank-correlated with some ranking scores, for a finite set of performances $\aSetOfPerformances$ sharing the same class priors.

\textbf{Step 1.}
Determine the priors $(\priorneg,\priorpos)$ such that $\aPerformance(\randVarGroundtruthClass=\classNeg)=\priorneg$ and $\aPerformance(\randVarGroundtruthClass=\classPos)=\priorpos$, $\forall\aPerformance\in\aSetOfPerformances$.

\textbf{Step 2.}
Determine some importance values $\randVarImportance_A$ and $\randVarImportance_B$ that correspond to the base scores. To do this, one can, for example, consult~\cref{tbl:ranking-scores}. This gives us the scores $\rankingScore[\randVarImportance_A]$ and $\rankingScore[\randVarImportance_B]$, which are perfectly rank-correlated with $\aScore_A$ and $\aScore_B$ for the class priors $(\priorneg,\priorpos)$.

\textbf{Step 3.}
Move the importance points along their iso-ranking line to place them on the plane $\itn \, \priorneg + \itp \, \priorpos = \ifp \, \priorneg + \ifn \, \priorpos$ (see \cref{fig:tetrahedron-ranking-scores}, right). For an importance $\randVarImportance$ (whether $\randVarImportance_A$ or $\randVarImportance_B$), calculate $\paramPriorsA(\randVarImportance)$ and $\paramPriorsB(\randVarImportance)$ using \cref{eq:paramPriors}, then apply \cref{eq:our-plane} to obtain $\randVarImportance'$. So we replace $\rankingScore[\randVarImportance_A]$ and $\rankingScore[\randVarImportance_B]$ respectively by $\rankingScore[\randVarImportance_A']$ and $\rankingScore[\randVarImportance_B']$ without changing the rankings. The optimal tradeoff between $\aScore_A$ and $\aScore_B$ is therefore the same as that between $\rankingScore[\randVarImportance_A']$ and $\rankingScore[\randVarImportance_B']$.

\textbf{Step 4.}
Consider the universal shortest path between $\rankingScore[\randVarImportance_A']$ and $\rankingScore[\randVarImportance_B']$, which is obtained by linearly interpolating the importances: $\randVarImportance=(1-w)\,\randVarImportance_A' + w\,\randVarImportance_B'$ with $w \in [0,1]$. In the plane $\itn \, \priorneg + \itp \, \priorpos = \ifp \, \priorneg + \ifn \, \priorpos$, this is a line segment. In the $(\paramPriorsA,\paramPriorsB)$ space, this is also a line segment.

\textbf{Step 5.}
Determine the mean generator $f$. Any score $\rankingScore$ on the shortest path is a $f$-mean of $\rankingScore[\randVarImportance_A']$ and $\rankingScore[\randVarImportance_B']$ generated by an homographic function $f:[0,1]\rightarrow\realNumbers:x\mapsto\frac{Ax+B}{Cx+D}$, whose coefficients are given by \cref{eq:homography-coefficients} and depend only on $\da=\paramPriorsA(\randVarImportance_B')-\paramPriorsA(\randVarImportance_A')$ and $\db=\paramPriorsB(\randVarImportance_B')-\paramPriorsB(\randVarImportance_A')$. We have $\rankingScore = f^{-1} \left[ (1-w) \, f \circ \rankingScore[\randVarImportance_A'] + w \, f \circ \rankingScore[\randVarImportance_B'] \right]$.

\textbf{Step 6.}
For each performance $\aPerformance\in\aSetOfPerformances$, precompute $(f \circ \rankingScore[\randVarImportance_A']) (\aPerformance)$ and $(f \circ \rankingScore[\randVarImportance_B']) (\aPerformance)$. These values will be used when \cref{eq:weight-for-boundry} will be applied in the next step.

\textbf{Step 7.}
Determine all the values of $w$ such that the universal shortest path crosses a boundary between neighboring rankings. To do this, apply \cref{eq:weight-for-boundry} for each pair $\aPerformance_i,\aPerformance_j\in\aSetOfPerformances$. Indeed, for $n=|\aSetOfPerformances|$ performances, this step requires  $O(n^2)$ uses of \cref{eq:weight-for-boundry} (this the same time complexity as one has with a naive implementation of Kendall's $\tau$). In practice, however, this is not a big deal: even for $n=1000$ (hardly met in practice), this takes less than a second on a modern laptop.

\textbf{Step 8.}
On the universal shortest path, choose a score $\rankingScore[\randVarImportance^*]$ that is equidistant from the two starting scores. To do this, determine $w^*$ using \cref{eq:closed-form-expression}. Note that $\distKendall(\rankingScore[\randVarImportance^*];\rankingScore[\randVarImportance_A']) = \distKendall(\rankingScore[\randVarImportance^*];\rankingScore[\randVarImportance_A]) = \distKendall(\rankingScore[\randVarImportance^*];\aScore_A)$ and $\distKendall(\rankingScore[\randVarImportance^*];\rankingScore[\randVarImportance_B']) = \distKendall(\rankingScore[\randVarImportance^*];\rankingScore[\randVarImportance_B]) = \distKendall(\rankingScore[\randVarImportance^*];\aScore_B)$. So, one has $\distKendall(\rankingScore[\randVarImportance^*];\rankingScore[\randVarImportance_A'])=\distKendall(\rankingScore[\randVarImportance^*];\rankingScore[\randVarImportance_B'])$ \iif $\distKendall(\rankingScore[\randVarImportance^*];\aScore_A)=\distKendall(\rankingScore[\randVarImportance^*];\aScore_B)$. When this equality is achieved, $\rankingScore[\randVarImportance^*]$ is a solution to \cref{eq:Karcher-mean}. It minimizes Fréchet variance and is thus a Karcher mean between $\aScore_A$ and $\aScore_B$, \wrt $\distKendall$.

\textbf{Step 9.}
Determine the parameters of the score $\rankingScore[\randVarImportance^*]$ that lies at this midpoint. This is an optimal tradeoff given by $\randVarImportance^*=(1-w^*) \, \randVarImportance_A' + w^* \, \randVarImportance_B'$. Any score that increases strictly with $\rankingScore[\randVarImportance^*]$ is also suitable. In particular, any ranking score $\rankingScore$ such that $\paramPriorsA(\randVarImportance)=\paramPriorsA(\randVarImportance^*)$ and $\paramPriorsB(\randVarImportance)=\paramPriorsB(\randVarImportance^*)$ or, equivalently, $\paramAllA(\randVarImportance)=\paramAllA(\randVarImportance^*)$ and $\paramAllB(\randVarImportance)=\paramAllB(\randVarImportance^*)$ is an optimal tradeoff between $\aScore_A$ and $\aScore_B$.

\section{Experiments}
\label{sec:experiments}

\subsection{Arbitrarily chosen means: a recipe for trouble}

\newcommand{\setOfPerformancesA}{\Pi_A}
\newcommand{\setOfPerformancesB}{\Pi_B}
\newcommand{\setOfPerformancesC}{\Pi_C}

\begin{figure}
    \centering
    \hfill
    \subfloat[With $\setOfPerformancesA$.]{
        \includegraphics[width=0.32\linewidth]{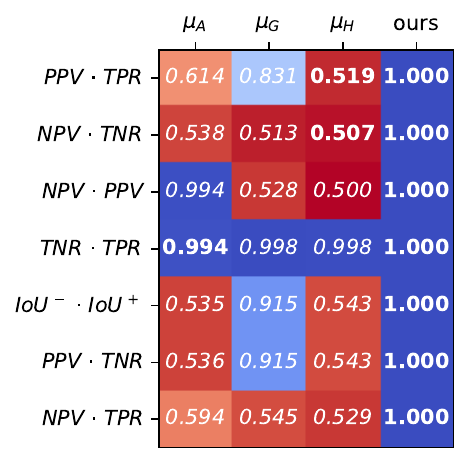}
    }
    \hfill
    \hfill
    \subfloat[With $\setOfPerformancesB$.]{
        \includegraphics[width=0.32\linewidth]{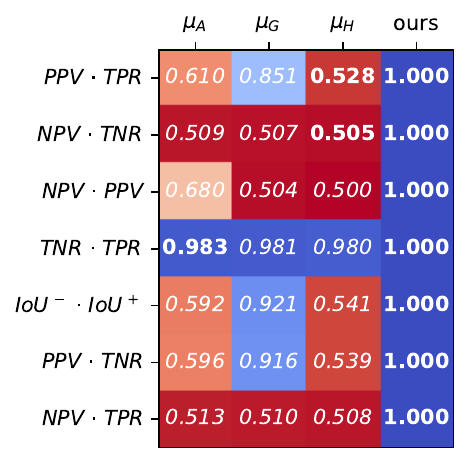}
    }
    \hfill
    \hfill
    \subfloat[With $\setOfPerformancesC$.]{
        \includegraphics[width=0.32\linewidth]{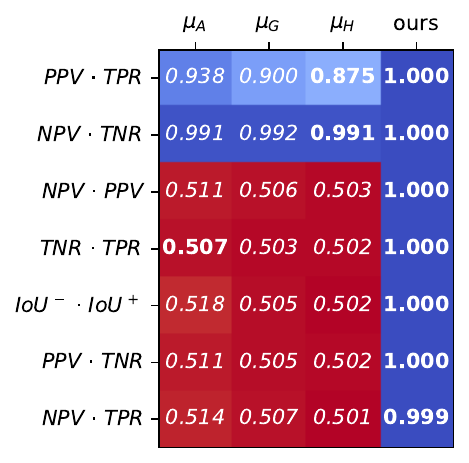}
    }
    \hfill
    \caption{Optimality $\optimality$ obtained with each of the three Pythagorean means ($\mu_A$: arithmetic, $\mu_G$: geometric, $\mu_H$: harmonic) and with our method, for seven pairs of scores and three sets of $1000$ performances corresponding to $\priorpos=0.01$. Our method always yields an optimal result ($\optimality=1$, in blue), but arbitrarily chosen means are only approximately optimal in a few specific cases. In fact, the optimality value is often barely above 0.5 (the worst, in red). The meaningful rankings (\ie, the tradeoffs satisfying the axioms of performance-based ranking \citep{Pierard2025Foundations}) are in bold.}
    \label{fig:experiment-1}
\end{figure}

The purpose of our \first experiment is to highlight the problems that arise when a tradeoff is made between two scores $(\scoreA,\scoreB)$ using an arbitrarily chosen mean $\mu$. We illustrate this using three common means (the arithmetic, the geometric, and the harmonic) for seven tradeoffs (three of them are tradeoffs between a score and its inverse), based on three large sets of performances. These sets consist of performances drawn at random, \iid,  following three performance distributions that have been studied theoretically in \citet{Pierard2026What} in the specific case of the $\scorePrecision$-$\scoreRecall$ tradeoff. In the case of the set $\setOfPerformancesA$, the distribution is uniform for fixed priors; this corresponds to a uniform distribution in $ROC=(\scoreFPR,\scoreTPR)\in[0,1]^2$. For $\setOfPerformancesB$, the distribution remains uniform with fixed priors but is restricted to performances above the no-skills (\ie, $\scoreTPR\ge\scoreFPR)$. Finally, for $\setOfPerformancesC$, we restrict even further and consider a uniform distribution over $(\scoreFPR,\scoreTPR)\in[0,\priorpos]\times[\priorpos,1]$.

For each pair of scores for which a compromise must be determined, we calculate the values taken by these two scores as well as their means, across all performances. Then, we calculate three Kendall correlations: between the two base scores, between the first score and the mean, and between the mean and the second score. From these values, we derive the optimality. It is defined in \citet{Pierard2026What} and can be generalized as:
$%
    \optimality(\aScore) 
    = 1 - \nicefrac{
        \left| \distKendall(\scoreA;\aScore) - \distKendall(\aScore;\scoreB) \right|
    }{
        2 \distKendall(\scoreA;\scoreB)
    }
$, %
where $\aScore$ is the proposed tradeoff. For a mean $\mu$, $\aScore=\mu(\scoreA;\scoreB)$. On any shortest path, the optimality increases linearly with Kendall's distance from both base scores (where it takes the minimal value $\optimality=\nicefrac12$) towards the optimal tradeoff (where it takes the maximal value $\optimality=1$). The optimality can be interpreted as the proportion of pairs of performances for which $\aScore$ and the optimal tradeoff do not agree, among those for which the ordering is not trivial (\ie, $\scoreA$ and $\scoreB$ do not agree). Indeed, the optimality can also be computed for any ranking score that our theory recommends.

The results of this first experiment are given in \cref{fig:experiment-1}. One can see that the ranking score recommended by our theory has an optimality of $1$, as expected, no matter the set of performances and the pair of scores that are considered. This is never the case with arbitrarily chosen means. With a prior $\priorpos=0.01$, the optimality is often close to $0.5$ (the worst scenario) with arbitrarily chosen means. In this figure, we have written in bold the values for the cases in which the tradeoff score satisfies the axioms of performance-based ranking. It is always the case with the tradeoffs we recommend as they are ranking scores. However, it is rarely the case with the classic means (for the readers interested to dive deeper into this topic, this is because the third axiom is not satisfied).

\subsection{Case studies: with reference distributions}

\definecolor{mplblue}{RGB}{31,119,180}
\definecolor{mplorange}{RGB}{255,127,14}
\definecolor{mplgreen}{RGB}{44,160,44}
\definecolor{mplred}{RGB}{214,39,40}
\definecolor{mplpurple}{RGB}{148,103,189}
\definecolor{mplbrown}{RGB}{140,86,75}
\definecolor{mplpink}{RGB}{227,119,194}

\newcommand{\NPVxTPR}{\textcolor{mplpink}{$\blacktriangle$}}
\newcommand{\TNRxPPV}{\textcolor{mplbrown}{$\blacktriangledown$}}
\newcommand{\TNRxNPV}{\textcolor{mplorange}{$\blacktriangleleft$}}
\newcommand{\PPVxTPR}{\textcolor{mplblue}{$\blacktriangleright$}}
\newcommand{\TNRxTPR}{\textcolor{mplred}{$\filleddiamond$}}
\newcommand{\NPVxPPV}{\textcolor{mplgreen}{$\blacksquare$}}
\newcommand{\IoUxIoU}{\textcolor{mplpurple}{\ding{54}}} %

\begin{figure}
    \centering
    \hfill
    \subfloat[$\setOfPerformancesA$: uniform\label{fig:experiment_2_tile_distri_a}]{
        \includegraphics[width=0.32\linewidth]{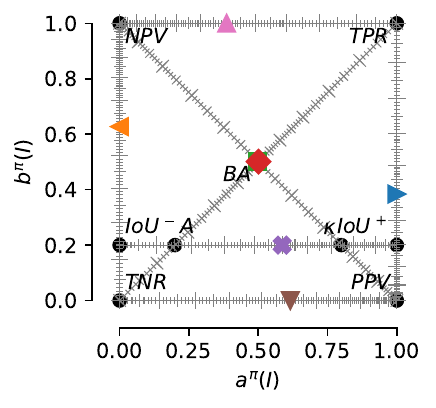}
    }
    \hfill
    \hfill
    \subfloat[$\setOfPerformancesB$: uniform, above no-skills\label{fig:experiment_2_tile_distri_b}]{
        \includegraphics[width=0.32\linewidth]{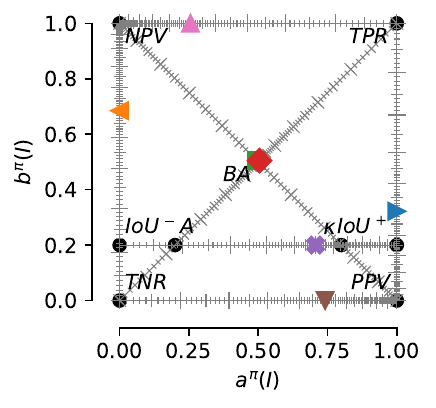}
    }
    \hfill
    \hfill
    \subfloat[$\setOfPerformancesC$: uniform , close to oracle\label{fig:experiment_2_tile_distri_c}]{
        \includegraphics[width=0.32\linewidth]{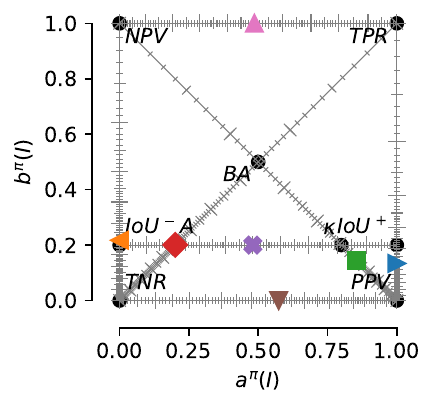}
    }
    \hfill
    \caption{
        Case studies with large performances sets randomly drawn from reference distributions with fixed priors ($\priorpos=0.2$). The colored markers indicate, in the $(\paramPriorsA,\paramPriorsB)$ space, the performance orderings that are optimal tradeoffs for the pairs 
        $\scorePPV \cdot \scoreTPR$~(\PPVxTPR), 
        $\scoreNPV \cdot \scoreTNR$~(\TNRxNPV), 
        $\scoreNPV \cdot \scorePPV$~(\NPVxPPV), 
        $\scoreTNR \cdot \scoreTPR$~(\TNRxTPR), 
        $\scoreIoUNeg \cdot \scoreIoUPos$~(\IoUxIoU),
        $\scorePPV \cdot \scoreTNR$~(\TNRxPPV), 
        and $\scoreNPV \cdot \scoreTPR$~(\NPVxTPR).
    }
    \label{fig:experiment-2}
\end{figure}

The goal of our \second experiment is to illustrate, within the $\ourSpace$ space, where the optimal tradeoffs lie for the seven pairs of base scores and for the three distributions of performances of the previous experiment. For each, we draw $1000$ performances to have large sets. Considering reference distributions is interesting because the organizer of a challenge is often required to announce the score that will be used to rank participants before receiving their actual performance results. For the organizer, one option is to choose a reference distribution and use it to select a score.%

In \cref{fig:experiment-2}, the rankings induced by the base scores, $\scoreAccuracy$, $\scoreBalancedAccuracy$, and $\scoreCohenKappa$ are shown with labeled black dots. Their locations can depend on the priors, but never on the set of performances. In the $\ourSpace$ space, the shortest path is the line segment connecting the base scores. The position of the optimal tradeoff on it depends on the set of performances. Along the seven studied segments, we draw graduated rulers such that the (not uniform) spacing between the marks is proportional to $\distKendall$. The optimal tradeoff (colored point) is thus at the median mark. The marks on the rulers allow to visualize how the optimal tradeoff would move if one weighted non-uniformly the two base scores in $\sigma^2(\randVarImportance)$ (see \cref{eq:Karcher-mean}). In this case, one needs to replace the median by a quantile in \cref{eq:closed-form-expression}.

We can see that the optimal tradeoff between precision ($\scorePPV$) and recall ($\scoreTPR$) is not $\scoreFOne$ (same ranking as $\scoreIoUPos$), but always corresponds to one of the $\scoreFBeta$ scores ($\paramPriorsA=1$ line). This confirms the results of \citet{Pierard2026What}. We also see that the behavior is, broadly speaking, quite similar between $\setOfPerformancesA$ and $\setOfPerformancesB$, but quite different with $\setOfPerformancesC$, particularly for the optimal tradeoff between $\scoreTNR$ and $\scoreTPR$ (close to the balanced accuracy $\scoreBalancedAccuracy$ for the first two sets; close to the accuracy $\scoreAccuracy$ for the third) and for the optimal tradeoff between $\scoreNPV$ and $\scorePPV$ (close to the balanced accuracy $\scoreBalancedAccuracy$ for the first two datasets; in the lower right quadrant of Cohen’s $\scoreCohenKappa$ for the third). There is generally no universal optimal tradeoff for a given pair of scores.

\subsection{Case studies: with real-world rankings}

\begin{figure}
    \centering
    \hfill
    \subfloat[With the $10$ performances reported in \citet{Bouwmans2026Illegal} for the \emph{Illegal Waste Dumping Detection} challenge, when the evaluation is done on the LS.\label{fig:experiment_3_results_contest_iwdd_2026_LS}]{
        \includegraphics[width=0.23\linewidth]{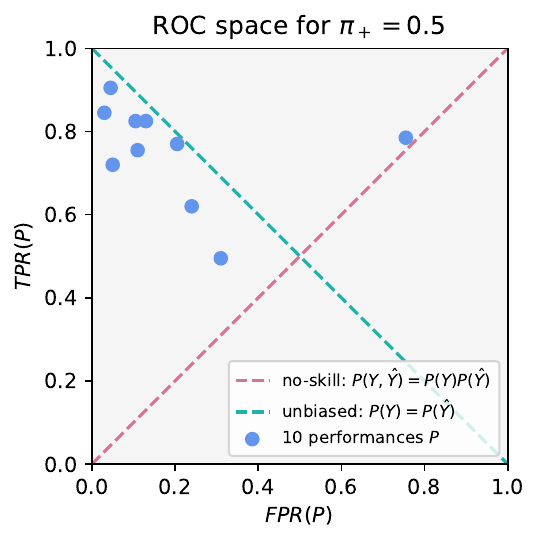}
        \includegraphics[width=0.23\linewidth]{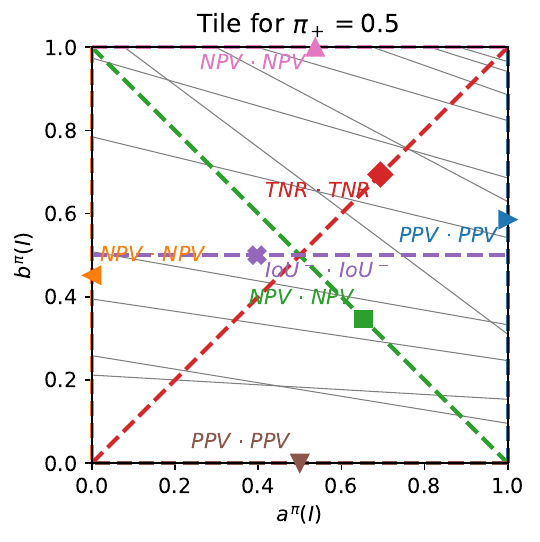}
    }
    \hfill
    \hfill
    \subfloat[With the $10$ performances reported in \citet{Bouwmans2026Illegal} for the \emph{Illegal Waste Dumping Detection} challenge, when the evaluation is done on the TS.\label{fig:experiment_3_results_contest_iwdd_2026_TS}]{
        \includegraphics[width=0.23\linewidth]{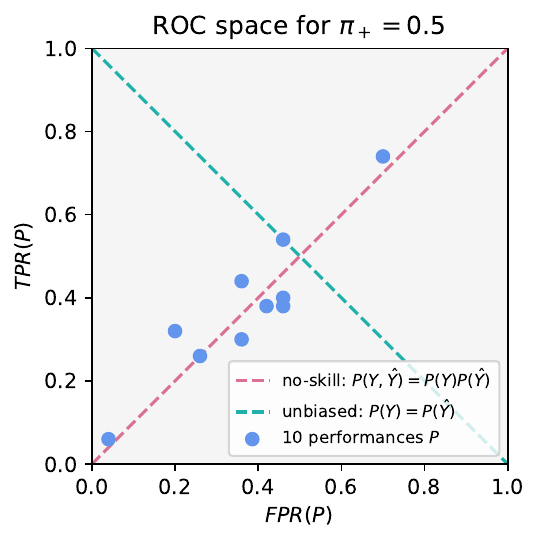}
        \includegraphics[width=0.23\linewidth]{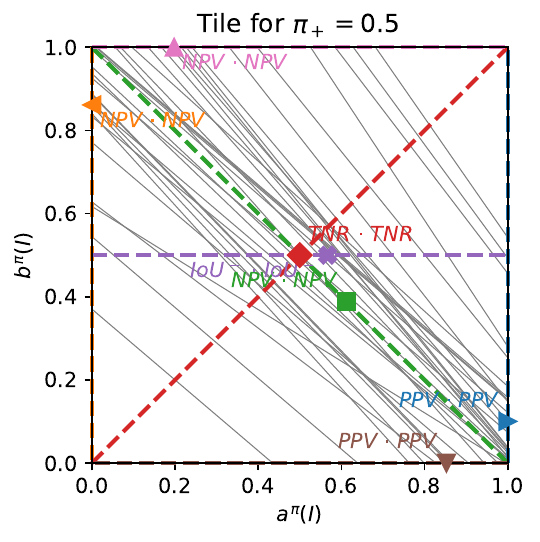}
    }
    \hfill
    
    \hfill
    \subfloat[With the performances given in \citet{Halin2024AHitchhikers-arxiv} for $74$ semantic segmentation models from \emph{MMSegmentation}~\citep{Mmsegmentation2020}.\label{fig:experiment_3_semantic_segmentation_methods}]{
        \includegraphics[width=0.23\linewidth]{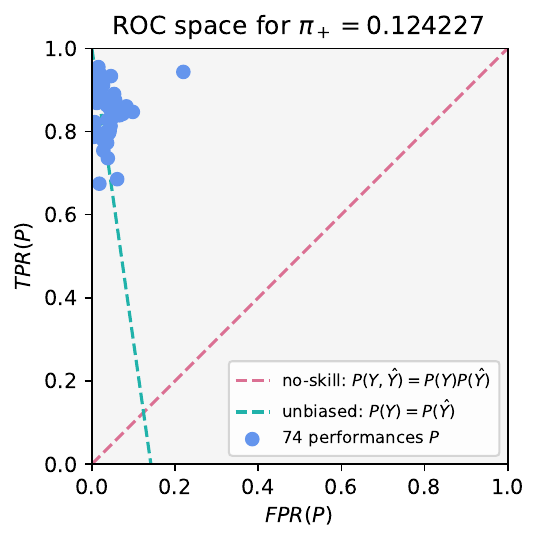}
        \includegraphics[width=0.23\linewidth]{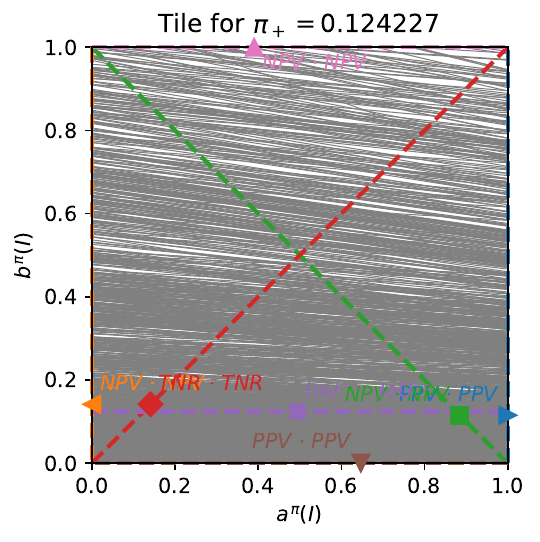}
    }
    \hfill
    \hfill
    \subfloat[With the performances of $57$ background subtraction methods reported on \href{https://changedetection.net}{changedetection.net} (CDnet 2014~\citep{Wang2014CDnet}) for the \emph{blizzard} video.\label{fig:experiment_3_cdnet_2014_34_blizzard}]{
        \includegraphics[width=0.23\linewidth]{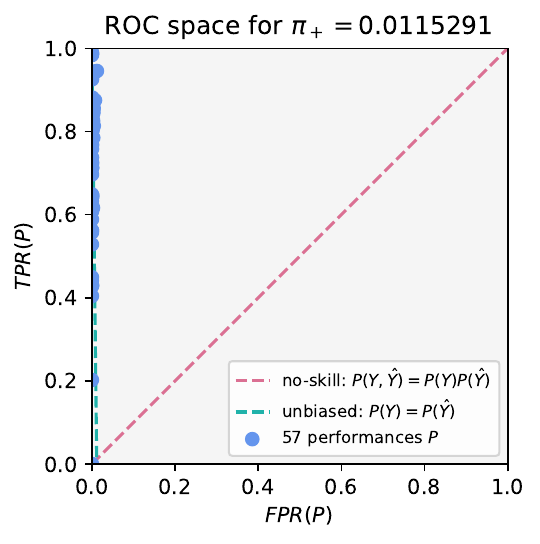}
        \includegraphics[width=0.23\linewidth]{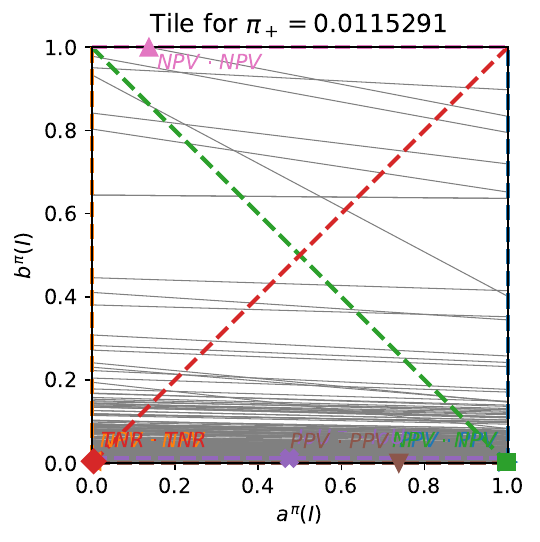}
    }
    \hfill
    \caption{Case studies with 4 real-world performance sets. For each of them, the performances are shown in the ROC %
    space, and the optimal tradeoffs for 7 pairs of scores are shown in $\ourSpace$.}
    \label{fig:experiment-3}
\end{figure}

The goal of our \third experiment is to show what happens with real-world performance sets whose size is significantly smaller than those we studied in the previous experiment. We considered four published or publicly available performance sets related to computer vision problems. The results are shown in \cref{fig:experiment-3}. There, we have plotted in the $\ourSpace$ space the frontiers between rankings. They correspond to the scores for which any two performances are considered equivalent. It turns out that, in the case of our parameterization adapted to the priors, $\rankingScore(\aPerformance_i)=\rankingScore(\aPerformance_j)$ holds if and only if the performance orderings lie on the line $\zeta_a \, \paramPriorsA(\randVarImportance) + \zeta_b \, \paramPriorsB(\randVarImportance) + \zeta = 0$ with $ \zeta_a = \scoreFPR(\aPerformance_i) \, \scoreFNR(\aPerformance_j) - \scoreFPR(\aPerformance_j) \, \scoreFNR(\aPerformance_i) $, $ \zeta_b = \scoreTPR(\aPerformance_i) \, \scoreTNR(\aPerformance_j) - \scoreTPR(\aPerformance_j) \, \scoreTNR(\aPerformance_i) $, and $ \zeta = \scoreTNR(\aPerformance_i) \, \scoreFPR(\aPerformance_j) - \scoreTNR(\aPerformance_j) \, \scoreFPR(\aPerformance_i) $. On any given shorter path, the number of times the path crosses one of these lines on either side of the optimal tradeoffs is equal.

The two first cases are based on the $10$ performances reported in \citet{Bouwmans2026Illegal} regarding the results of the Illegal Waste Dumping Detection (IWDD) challenge 2026. The authors reported the performances measured on the learning set (LS) and on the test set (TS) to highlight the discrepancy between the two (overfitting).
Both LS and TS are balanced ($\priorneg=\priorpos$).%
We examine the LS case in \cref{fig:experiment_3_results_contest_iwdd_2026_LS} and the TS case in \cref{fig:experiment_3_results_contest_iwdd_2026_TS}. In the latter case, all performances are close to the no-skills (rising diagonal of ROC, \ie $\randVarGroundtruthClass \indep_\aPerformance \randVarPredictedClass$), and the optimal tradeoffs approach the descending diagonal of $\ourSpace$ where the no-skill performances are put on an equal footing. In the extreme case where all compared performances were no-skill, all the boundaries between the rankings would coincide with the descending diagonal, and consequently, all optimal tradeoffs would lie on it.

The last two cases are for the performances reported by \citet{Halin2024AHitchhikers-arxiv} for $74$ semantic segmentation models from \emph{MMSegmentation}~\citep{Mmsegmentation2020}, evaluated on BDD100K~\citep{Yu2020BDD100K} after grouping the semantic labels into two classes, %
and the performances of $57$ background subtraction methods (classifying pixels as background $\classNeg$ or foreground $\classPos$; for more information, see \citet{Bouwmans2019Deep,Bouwmans2019VisualHuman,Bouwmans2019VisualNatural}) measured on the video ``blizzard'' \citep{Goyette2012Changedetection,Wang2014CDnet}. They correspond, respectively, to $\priorpos \simeq 0.124$ and $\priorpos \simeq 0.012$. The comparison with \cref{fig:experiment_2_tile_distri_c} shows that the optimal tradeoffs are close to those predicted by the uniform distribution of performances, close to the oracle ($\setOfPerformancesC$).

\section{Conclusion}

Two-class classification is, conceptually, a simple task. However, the evaluation of the classifiers, as well as the analysis and comparison of their performances, remains relatively poorly understood. We have presented theoretical contributions that address the question: ``Which scores correspond to the optimal tradeoff between the rankings induced by two arbitrarily chosen scores?'' We did so in the specific but very common case of two-class classification with fixed priors, deliberately restricting ourselves to the family of ranking scores known to yield meaningful rankings (\ie, rankings interpretable from a performance perspective). We first presented the theory, then presented the results of three experiments demonstrating the practical applicability of the proposed technique.

\subsubsection*{Acknowledgments}
S. Pi{\'e}rard is funded by grant 2010235 (ARIAC by \href{https://www.digitalwallonia.be/en/}{DigitalWallonia4.ai}) of the SPW EER, Wallonia, Belgium; A. Deli{\`e}ge is a \href{https://www.frs-fnrs.be}{F.R.S.-FNRS} postdoc researcher.


\end{document}